\documentclass[oldversion]{aa}
\usepackage{txfonts}
\usepackage{graphicx}
\usepackage{natbib}
\bibpunct{(}{)}{;}{a}{}{,}

\def\intgr{{\it INTEGRAL}}
\def\bps{{\it BeppoSAX}}

\begin{document}
\title{\emph{INTEGRAL} discovery of non-thermal hard X-ray emission from the Ophiuchus cluster}
\author{D. Eckert\inst{1}, N. Produit\inst{1}, S. Paltani\inst{1}, A. Neronov\inst{1} \& T. J.-L. Courvoisier\inst{1}}
\offprints{Dominique Eckert, \email{Dominique.Eckert@obs.unige.ch}}

\institute{ISDC, Geneva Observatory, University of Geneva, 16, ch. d'Ecogia, CH-1290 Versoix, Switzerland}
\date{Received 15 October 2007 / Accepted 5 December 2007}

\abstract{We present the results of deep observations of the Ophiuchus cluster of galaxies with \emph{INTEGRAL} in the 3-80 keV band. We analyse 3 Ms of \emph{INTEGRAL} data on the Ophiuchus cluster with the IBIS/ISGRI hard X-ray imager and the JEM-X X-ray monitor. In the X-ray band using JEM-X, we show that the source is extended, and that the morphology is compatible with the results found by previous missions. Above 20 keV, we show that the size of the source is slightly larger than the PSF of the instrument, and is consistent with the soft X-ray morphology found with JEM-X and ASCA. Thanks to the constraints on the temperature provided by JEM-X, we show that the spectrum of the cluster is not well fitted by a single-temperature thermal Bremsstrahlung model, and that another spectral component is needed to explain the high energy data. We detect the high energy tail with a higher detection significance (6.4$\sigma$) than the \emph{BeppoSAX} claim $(2\sigma)$. Because of the imaging capabilities of JEM-X and ISGRI, we are able to exclude the possibility that the excess emission comes from very hot regions or absorbed AGN, which proves that the excess emission is indeed of non-thermal origin. Using the available radio data together with the non-thermal hard X-ray flux, we estimate a magnetic field $B\sim0.1-0.2\,\mu G$.}

\authorrunning{Eckert D. et al.}
\titlerunning{\emph{INTEGRAL} discovery of non-thermal hard X-ray emission from the Ophiuchus cluster}

\maketitle

\section{Introduction}

Clusters of galaxies are the biggest bound structures of the universe, and, according to the hierarchical scenario of structure formation, the latest ones to form. They are filled by a hot ($10^7-10^8$ K) plasma, called intra-cluster medium (ICM), and thus radiate in soft X-ray bands through thermal Bremsstrahlung. 

The most massive clusters should form by merging of smaller clusters, that would create shock waves when the ICM of the clusters merges \citep{sarazin}. Such events are probably the major source of heating in the ICM. Models predict that a large population of relativistic electrons should be created during the merging event, which are expected to radiate through synchrotron emission in the radio domain, and through inverse-Compton (IC) scattering with the cosmic microwave background (CMB) in the hard X-ray band \citep{ensslin}. Another possible model involves a population of multi-TeV electrons that would radiate in hard X-rays through synchrotron emission \citep{timokhin}. While the extended radio emission from several clusters has been known for a long time (see e.g. \citet{feretti}), there is still no firm detection of the hard X-ray emission. \emph{BeppoSAX} observations report on the detection of a hard tail in the X-ray spectrum of at least Coma \citep{fusco} and Abell 2256 \citep{fusco2256}, but these detections are rather weak and controversial \citep{rossetti}. Hence, confirmation of these results by other instruments is important.

The Ophiuchus cluster is the second brightest cluster in the 2-10 keV band. It is a nearby (z=0.028, \citet{johnston}) rich cluster located in the direction of the Galactic Center ($l=0.5^\circ, b=9.4^\circ$), with a very high plasma temperature ($kT\sim10$ keV). Using \emph{ASCA} data, \citet{watanabe} showed that the cluster is not dynamically relaxed and exhibits some regions with very high temperature ($kT>13$ keV). Because these characteristics are very similar to the Coma cluster, they concluded that this cluster has also experienced a major merging event in the recent past. In the radio domain, \citet{johnston} claimed that the source is associated with the steep-spectrum radio source MSH 17-203. The identification of this radio source as a radio halo from the Ophiuchus cluster implies the presence of relativistic electrons, and hence predicts the presence of a non-thermal high-energy tail in the X-ray spectrum. \citet{nevalainen} searched for such an excess in a sample of nearby clusters with \emph{BeppoSAX}/PDS. For the Ophiuchus cluster, their analysis reveals a mean temperature of $9.1\pm0.6$ keV, and a $2\sigma$ excess at hard X-rays. However, at such a low significance level, the excess might as well be due to statistical fluctuations or instrument systematic effects. Moreover, the PDS instrument on board \emph{BeppoSAX} was non-imaging, so it was impossible to extract any information on the morphology of the hard X-ray emission. Therefore, the spectrum might be contaminated by absorbed point sources.

In this paper, we present the results of deep (3 Ms) observations of the Ophiuchus cluster with the IBIS/ISGRI and JEM-X instruments on board \intgr, in the aim of investigating the presence of a high-energy tail. We present the specific method we have used to analyse ISGRI data. Finally, we use radio data to consider possible models for the non-thermal emission of this object.

\section{Data Analysis}
\label{secdata}
\subsection{ISGRI data analysis}
\label{secisg}
The IBIS/ISGRI instrument on board \emph{INTEGRAL} \citep{lebrun} is a wide-field ($29^\circ\times29^\circ$) coded-mask instrument sensitive in the 15-400 keV band. Its angular resolution (12 arcmin FWHM nominal) is of the same order of magnitude as the size of the core of the cluster (diameter $\sim 10'$, \citet{watanabe}). Because of the large field-of-view (FOV) and the fact that \emph{INTEGRAL} spends a significant part of its observing time in the Galactic Bulge region, the amount of data accumulated by the instrument on the Ophiuchus cluster is quite large, which allows us to reach a high signal-to-noise ratio for the cluster in the 17-60 keV band. Our analysis covers 1580 pointings (or Science Windows, ScWs), including 1493 ScWs from public data and 87 from the \emph{INTEGRAL} AO-4 Key Programme, for a total observing time of 3 Ms. However, the location of the source ($9.3^\circ$ from the Galactic Center and $14.5^\circ$ from the very bright X-ray binary Sco X-1) makes difficult the imaging analysis of this region with ISGRI. Furthermore, the periodic shape of the IBIS mask causes the images generated by the standard Offline Scientific Analysis software (OSA, \citet{cour}) to present ghosts of bright sources which are not completely removed at some specific positions in the deconvolved sky images. This adds important systematic errors to long exposure mosaics in the vicinity of bright sources.

In a coded-mask instrument, the sky images are produced by the deconvolution of the shadow patterns cast by the sources in the FOV on the plane of the sky. When observing a single source, a large fraction of the detector ($>45\%$) is not illuminated by the source, which allows us to measure the level of the background in each pointing. The detector image, or "shadowgram", produced when observing a field containing $n$ sources is a superposition of the shadow patterns of all the individual sources, 
\begin{equation}S(x,y)=\sum_{i=1}^n f_i\cdot PIF_i(x,y)+bB(x,y),\label{pif}\end{equation}
where $f_i$ and $PIF_i(x,y)$ are respectively the flux of the $i$th source and the shadow pattern (called "pixel illumination fraction") it casts, and $B(x,y)$ is a "background map", i.e. a model of the background, in the pixel with coordinates $(x,y)$. Therefore, $bB(x,y)$ gives the total background model in the pixel $(x,y)$. If the background was flat over the whole detector, the deconvolution process would not be sensitive to the background. Therefore, background maps take the deviations from the flat background into account. The properties of the coded mask method allow us to compute a background map by averaging the count rates of a pixel for many pointings, excluding the pixels illuminated by a strong source. To take the variability of the background into account, we computed a specific background map for each revolution of \intgr.

To the present day, the exact shape of the IBIS mask is not completely understood, so the PIFs created by the standard OSA software are not perfectly correct. The occurence of even a handful of wrongly modelled pixels can very strongly affect image reconstruction. The identification of these pixels is made difficult by the fact that bright sources do also affect counts in exposed pixels, which should not be removed. In order to identify badly modelled pixels, we procede in the following way: we fit the model described in Eq. (\ref{pif}) to the shadowgram and extract the background level and the flux of the sources present in the FOV. Then we remove bright sources and background from the shadowgram:
\begin{equation}S'(x,y)=S(x,y)-\sum_{i=1}^n f_i\cdot PIF_i(x,y)-bB(x,y),\end{equation}
with error map given by error propagation. In the case of a perfect PIF, the pixel distribution of the transformed detector image would be Gaussian, and hence 99.7\% of the pixels would have a value within $[\mu-3\sigma,\mu+3\sigma]$. Since the ISGRI detector contains $\sim15'000$ pixels, statistically we expect $\sim45$ of them to show a deviation from the mean value larger than $3\sigma$. However, we notice that $\sim150$ pixels deviate from the mean value by more than $3\sigma$, so we conclude that a large fraction of these pixels are incorrectly modelled. Then we perform a standard OSA 7.0 analysis ignoring these pixels.  We also neglect the outer parts of the image (with off-axis angle $>14^\circ$), where the systematic effects are the most important. Finally, we make a mosaic from the modified images in a standard way.

Unlike the image reconstruction process, the standard OSA spectral extraction procedure does not suffer from these problems as long as the input source catalog contains all sources brighter or comparable with the analysed source. Therefore, for spectral extraction we performed a standard OSA 7.0 analysis with a source catalog extracted from the mosaic image.

\subsection{JEM-X data analysis}

The JEM-X X-ray monitor on board \intgr\ \citep{lund} consists of two identical X-ray detectors with coded mask, JEM-X 1 and JEM-X 2, sensitive in the 3-35 keV band. It is designed for the spectroscopic and imaging study of the sources detected by IBIS, with a better spatial resolution (3.35 arcmin FWHM nominal) and a field-of-view of $7.5^\circ$ half-response in diameter. On the Ophiuchus cluster, we reached a 160 ks effective exposure time, for a total observing time of 350 ks.

The analysis of the Ophiuchus field with JEM-X does not present the same difficulties as those affecting the IBIS observations. Indeed, both the Galactic Center and Sco X-1 are outside the field of view of JEM-X. We therefore performed a completely standard image analysis, taking advantage of the improvements in the image reconstruction process introduced in OSA 7.0, in particular regarding the astrometry. In order to reach the deepest sensitivity of JEM-X, we combined all JEM-X 1 and JEM-X 2 ScWs in a single mosaic. Spectral extraction has been performed by extracting the fluxes from this mosaic in four separate energy bands chosen to provide clear detection in each individual band and to complement the ISGRI spectrum without gap. The fact that the Ophiuchus cluster is probably extended with JEM-X is not a problem, since the flux is integrated over the source. Therefore we simply let the width of the fitted Gaussian free. 

When extracting fluxes from JEM-X mosaics, there is no standard way in OSA to provide associated responses. The redistribution matrices (RMFs) of both JEM-X instruments are stable with time and quite close for both instruments; we therefore simply use the RMF of JEM-X 1. The area responses (ARFs) are however varying both with time and with the instrument. In particular, the low-energy efficiency changed very strongly (by a factor close to 3 for JEM-X 1) when the event rejection criteria were modified. It is therefore necessary to build an ARF suitable for the analysis of the spectrum extracted from the combined mosaic. To do this, we extract the correct ARF for each ScW and calculate the average of these ARFs weighted by the effective exposure time for each ScW, taking into account the dead time and the vignetting. As most observations use JEM-X 1 with the new rejection criterion, in practice the resulting ARF is quite close to that valid for any of the late JEM-X 1 ScWs. Therefore, the choice of a particular ARF does not affect qualitatively any of the results presented in this paper.

\section{Imaging results}
\label{secima}

\begin{figure}
\resizebox{\hsize}{!}{\hbox{
\fbox{\includegraphics{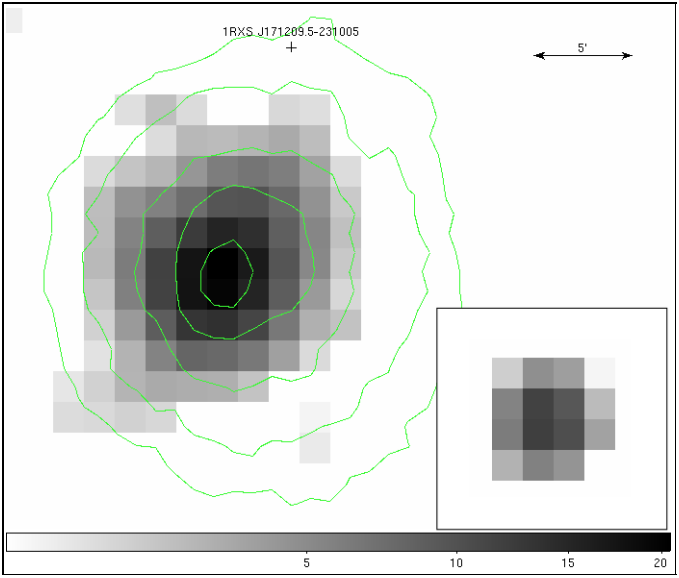}}}}
\caption{JEM-X significance image in the 3-18 keV band, with surface brightness contours from ASCA overlayed. The inset shows the image of a known point source in the same field, V 2400 Oph, to show the extended nature of the source. The cross shows the position of 1RXS J171209.5-231005, the nearest X-ray point source.}
\label{jemxima}
\end{figure}

As demonstrated in \citet{eckert} for the Coma cluster, the angular resolution of \intgr\ can be sufficient to get important information on the morphology of the hard X-ray emission from clusters. Since the Ophiuchus cluster is a very bright cluster below 10 keV, one can use the JEM-X monitor to study the change in morphology with energy. Fig. \ref{jemxima} shows the JEM-X mosaic image of the cluster in the 3-18 keV band, with contours from \emph{ASCA}/GIS observations of the cluster. For comparison, the inset shows the JEM-X image of the cataclysmic variable V 2400 Oph, in the same field. The extended nature of the Ophiuchus cluster in the JEM-X images is clear.

\begin{figure}
\resizebox{\hsize}{!}{\hbox{
\fbox{\includegraphics{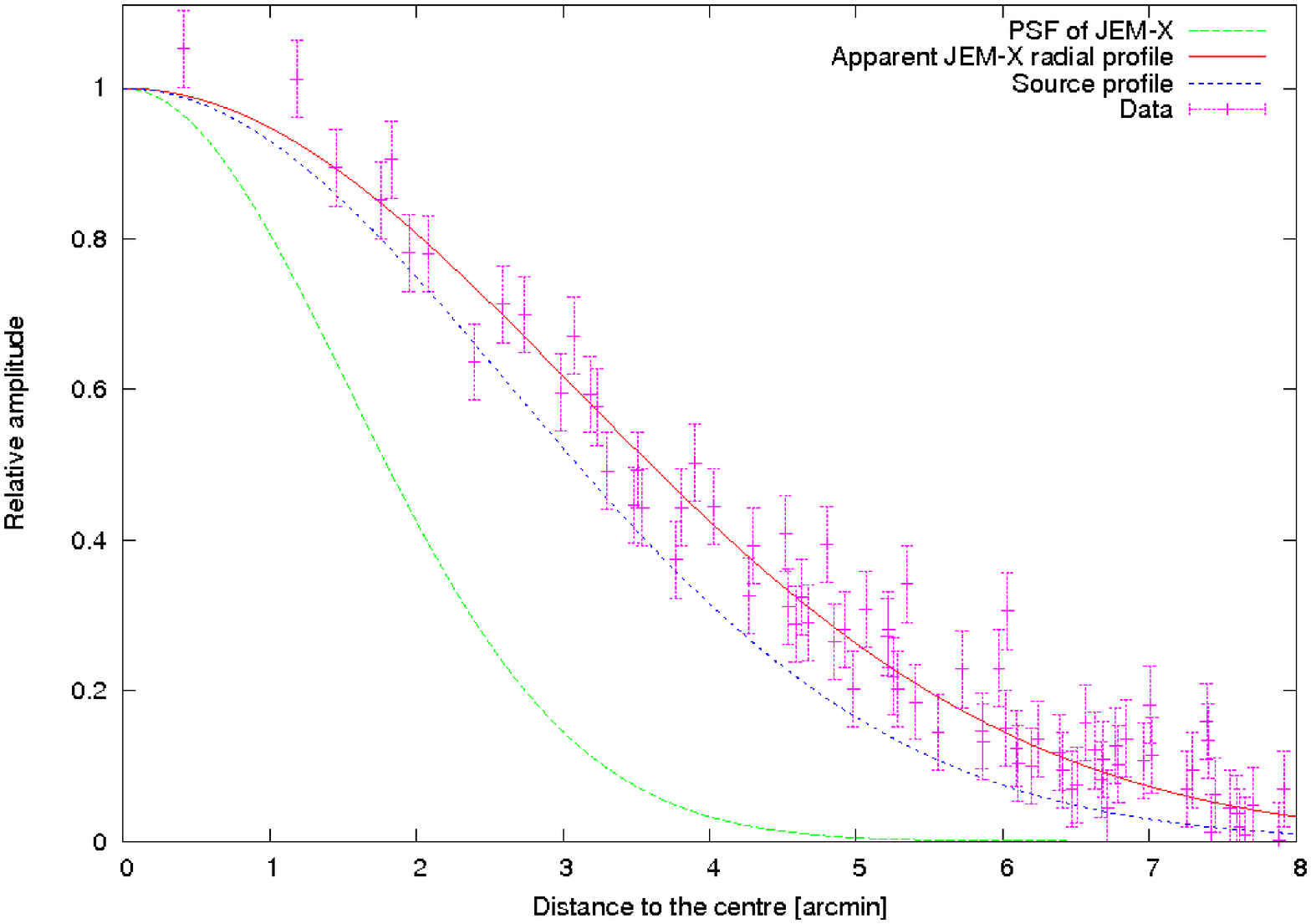}}}}
\caption{Radial profile of the source with a core radius of $3.1'$ extracted from JEM-X data (dashed blue line), compared to the PSF of the instrument (dashed green). The solid red line shows the total apparent JEM-X profile. The purple bars show the data points extracted from the JEM-X image.}
\label{rad}
\end{figure}

Because we detect only the core of the cluster, it is not possible to fit the surface brightness with an isothermal beta profile, but fitting the image with a Gaussian model,
\begin{equation}I(r)=A\exp\left(-\ln(2)\frac{r^2}{R^2}\right),\label{profile}\end{equation}
where $R$ is the half-width at half-maximum (HWHM), one can get an indication of the size of the detected region and of possible deviations from the spherical model. Fitting the JEM-X image shown in Fig. \ref{jemxima} with the model described in Eq. (\ref{profile}), we find $R=3.6\pm0.1$ arcmin. This value is significantly larger than the PSF of JEM-X (1.8 arcmin HWHM), which confirms that the source is seen as extended. This is, after SN 1006 \citep{kalemci}, the second report of an extended source for JEM-X. The $R$ parameter corresponds to the superposition of the radial profile of the source and of the PSF of the instrument, $R^2=R_{\mbox{\tiny{source}}}^2+HWHM_{\mbox{\tiny{JEM-X}}}^2$. Therefore, the angular size of the source at half-maximum becomes $R_{\mbox{\tiny{source}}}=3.1\pm0.1$ arcmin. Figure \ref{rad} shows the radial profile of the source compared to the PSF of the instrument, as well as the total apparent radial profile from JEM-X data. We can see that the radial profile is well modelled by a Gaussian. The extension of the source is clear.

\begin{figure}
\begin{center}
\resizebox{\hsize}{!}{\hbox{
\fbox{\includegraphics[height=6cm,width=6cm]{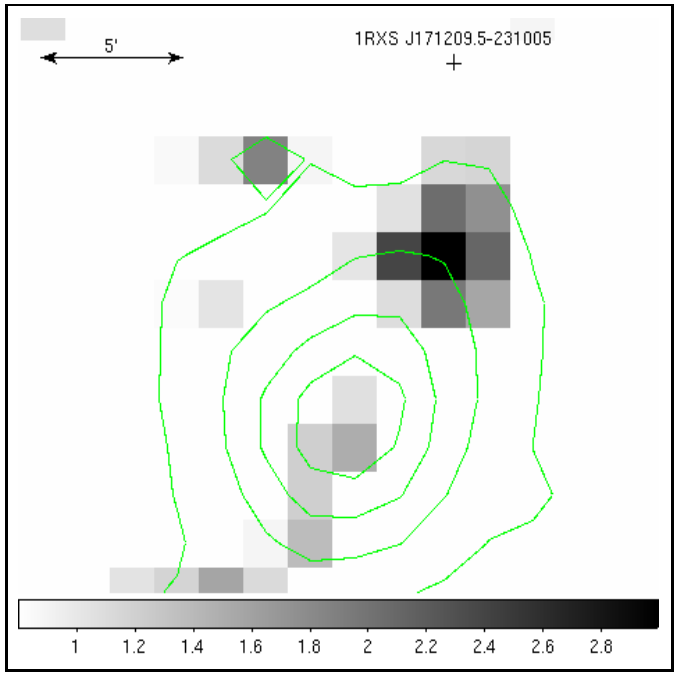}}}}
\end{center}
\caption{Residuals from the spherically symmetric model with JEM-X brightness contours overlaid, in units of $\sigma$. Similar to previous \emph{ASCA} observations, a significant excess is found North-West of the center of the cluster.}
\label{spher}
\end{figure}

From \emph{ASCA} data, \citet{watanabe} find the same value for the core of the cluster, $R_c=3.2'$. They also report on an extended region North-West of the center of the cluster that shows significant deviations from the spherical model. The authors note that this excess coincides with the closest X-ray point source, 1RXS J171209.5-231005. Figure \ref{spher} shows the residuals of the JEM-X image compared to the spherically symmetric model. A clear excess North-West of the center is also visible. However, it is clear that this excess is not due to 1RXS J171209.5-231005, which is located outside the region detected by JEM-X. The deviations from the spherical model are hence more likely due to over-densities or excess temperatures in the ICM, which indicates that the cluster is not dynamically relaxed, probably because of a merger in its recent history.

\begin{figure}
\begin{center}
\resizebox{\hsize}{!}{\vbox{
\hspace{-0.5cm}\includegraphics[height=6cm,width=9cm]{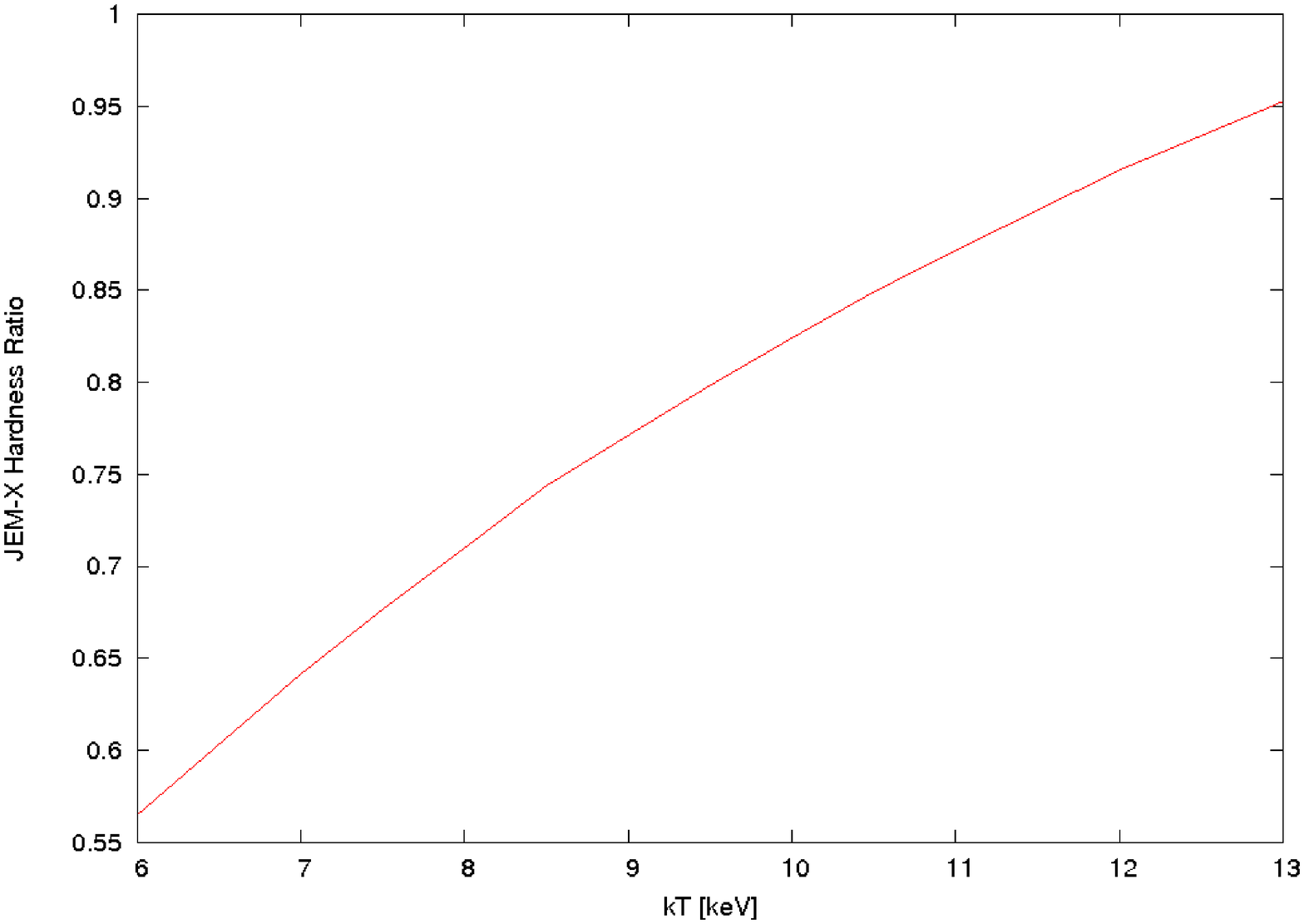}
\fbox{\includegraphics[height=6cm,width=6cm]{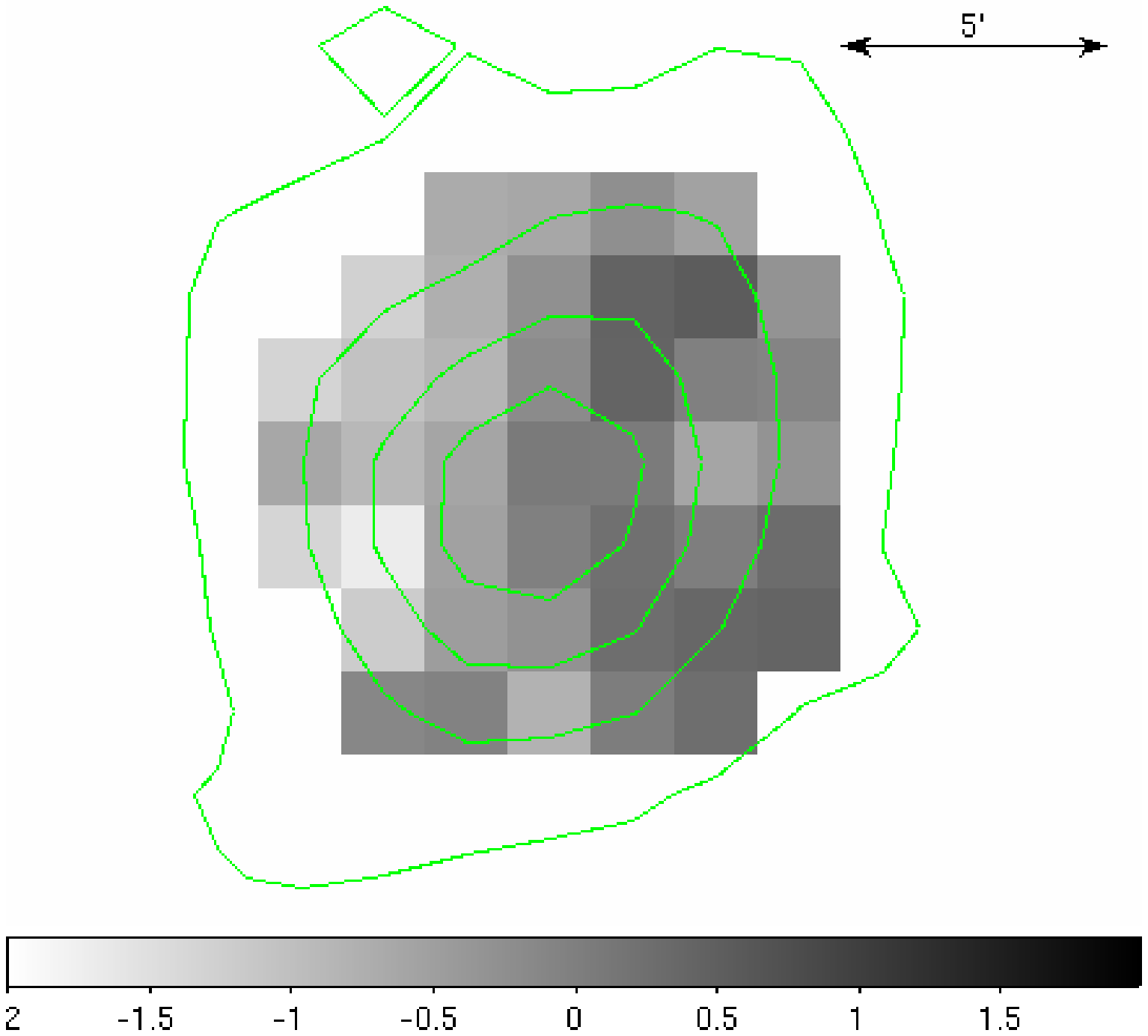}}}}
\end{center}
\caption{\emph{Top:} Simulated JEM-X 7-18/3-7 keV hardness ratio as a function of temperature. \emph{Bottom:} Deviations of the JEM-X 7-18/3-7 keV hardness ratio compared to the central value in units of $\sigma$, with JEM-X brightness contours overlaid. The map does not show any significant spatial dependence of the temperature within the detected region.}
\label{hard}
\end{figure}

We used the available JEM-X data to create a hardness ratio map (7-18 keV to 3-7 keV) in order to estimate the temperature variations within the detected region of the cluster. To estimate the dependence of the JEM-X 7-18 to 3-7 keV hardness ratio with temperature, we simulated JEM-X spectra for models with different input temperatures and computed the hardness ratios. The result is shown in the top panel of Fig. \ref{hard}. Because of the sensitivity up to 18 keV, the temperature dependence is steeper than in the case of \emph{ASCA}. In the center of the cluster, we find a hardness ratio $HR=0.78\pm0.09$, which corresponds to a temperature $kT=9.1_{-0.9}^{+1.4}$. The bottom panel of Fig. \ref{hard} shows the deviations of the hardness ratio compared to the central value, in units of $\sigma$. We find no deviations to the central value at significance level above 0.7$\sigma$, which corresponds to variations of $\pm 2$ keV compared to the central value. In \emph{ASCA} data, although there is evidence for a strong spatial dependence of the temperature, this dependence is found only in the outer parts of the cluster, while the core, which is the only region detected by JEM-X, shows an almost iso-thermal profile, compatible with the JEM-X results. This analysis shows that a single-temperature model should describe well the total JEM-X spectrum.

\begin{figure}
\resizebox{\hsize}{!}{\fbox{\hbox{
\includegraphics{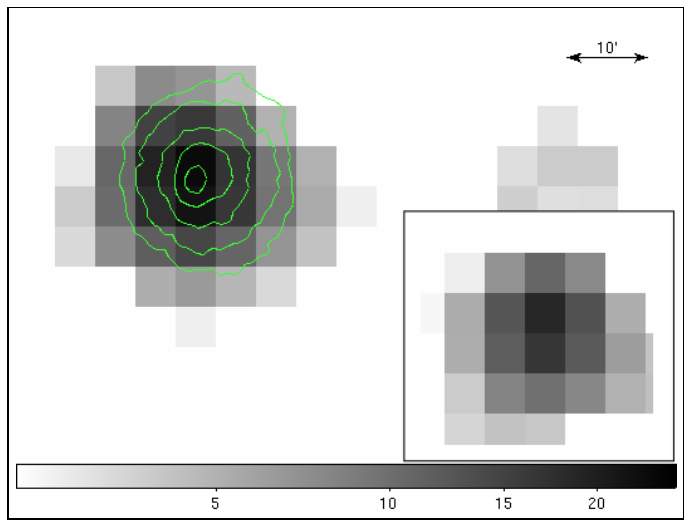}}}}
\caption{ISGRI significance map in the 20-40 keV band, with surface brightness contours from ASCA in green. The inset shows the ISGRI image of a point source, the cataclysmic variable V 2400 Oph, for comparison.}
\label{isgrima}
\end{figure}

Fig. \ref{isgrima} shows the ISGRI significance image in the 20-40 keV band extracted from the mosaic created with the method described in Sect. \ref{secisg}. Because of the lower spatial resolution, the extension of the source is not obvious (see the image of V 2400 Oph in the inset for comparison). However, a more detailed analysis of the source reveals that the source is indeed extended also for ISGRI. Fitting the image with the model described in Eq. (\ref{profile}), one finds $R=8.9\pm0.3$ arcmin, which is slightly larger than the PSF of the instrument (7.2 arcmin HWHM). Taking into account the contribution of the radial profile of the source and of the PSF of the instrument similar to the analysis presented above for JEM-X, this corresponds to $R_{source}=5.1_{-0.6}^{+0.5}$. Even though the source is seen as extended, the apparent size of the Ophiuchus cluster is smaller than the PSF of the instrument. To estimate the error on the standard flux extraction in the case of extended sources, \citet{renaud} presented the results of simulations assuming a uniform disk. For an angular size of 5', they evaluate the relative error to be $\sim5\%$. Since the cluster has a peaked radial profile instead of a uniform disk, the relative error on the flux extracted by the OSA spectral extraction tool is $<5\%$. For this reason, from now on we will treat the source as point-like for ISGRI. 

\begin{figure}[!tbh]
\resizebox{\hsize}{!}{\fbox{\hbox{
\includegraphics{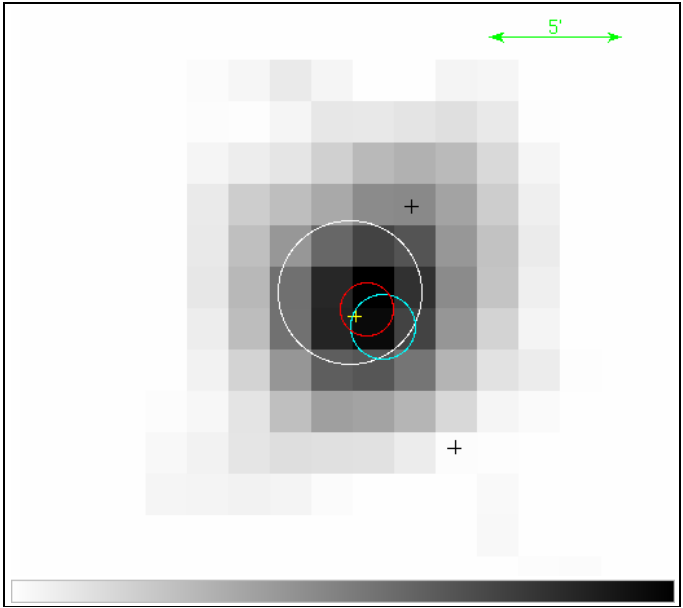}}}}
\caption{90\% error circles for the position of the center of the source in the 20-24 (red), 24-30 (blue) and 30-40 keV (white) ISGRI mosaic images, overlaid on the JEM-X 3-18 keV image. The yellow cross shows the position of the centroid of the cluster from the \emph{Chandra} image. The black crosses show the position of the two brightest X-ray point sources detected in the \emph{Chandra} image.}
\label{figpos}
\end{figure}

Since the apparent size of the source is smaller than the PSF of ISGRI, it is not possible to study in detail the hard X-ray morphology of the source. However, it is possible to measure the best fit position for the center of the source in different energy bands. Figure \ref{figpos} shows the 90\% error circles for the position of the source in the 20-24, 24-30 and 30-40 keV bands overlaid on the JEM-X 3-18 keV mosaic image. The yellow cross shows the position of the center of the cluster from the \emph{Chandra} image. The black crosses represent the position of the 2 brightest point sources close to the center in the \emph{Chandra} image. We can see on this image that the hard X-ray emission is compatible with the soft X-ray morphology of the cluster, and is not displaced towards any of the weak point sources seen in the \emph{Chandra} image, which excludes the possibility that these point sources contribute in a significant way to the high-energy spectrum.

\section{Spectral analysis}
\label{secspec}

\subsection{\intgr\ broad band spectrum of the cluster}

Since, unlike the case of the Coma cluster, the source is only very slightly larger than the PSF of ISGRI, we can use the standard OSA 7.0 spectral extraction tool with our time-dependent background maps to get an accurate flux estimation. Starting from OSA version 7.0, the calibration of ISGRI is now valid down to 17 keV. Because of the large exposure time, we were able to reach a high signal-to-noise ratio in both JEM-X (3-18 keV band) and ISGRI (17-60 keV band). In the 40-60 keV band, the detection significance reaches $6\sigma$. In the 60-80 keV band, the source is marginally detected by ISGRI at the $1.5\sigma$ level. It is important to note that no cross-calibration factor between ISGRI and JEM-X data was needed for this work. Introducing such a factor in the XSPEC fitting model results in a cross-calibration of 1.0, which shows the excellent calibration work done by the ISGRI and JEM-X teams for the release of OSA 7.0.

\begin{figure}
\resizebox{\hsize}{!}{\hbox{
\includegraphics[angle=270]{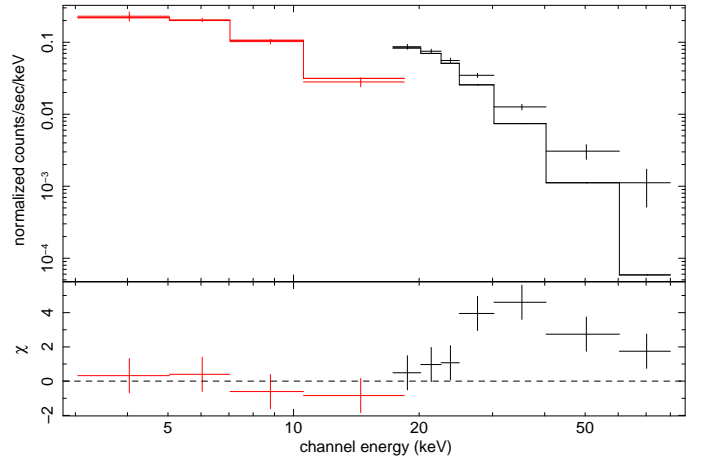}}}
\caption{JEM-X/ISGRI combined spectrum in the 3-80 keV band. The solid line is a fit to the 3-20 keV part of the spectrum with a single-temperature MEKAL model. The bottom panel shows the residuals from the model. The hard X-ray excess is clear.}
\label{spec1}
\end{figure}

\begin{figure}
\resizebox{\hsize}{!}{\hbox{
\includegraphics[angle=270]{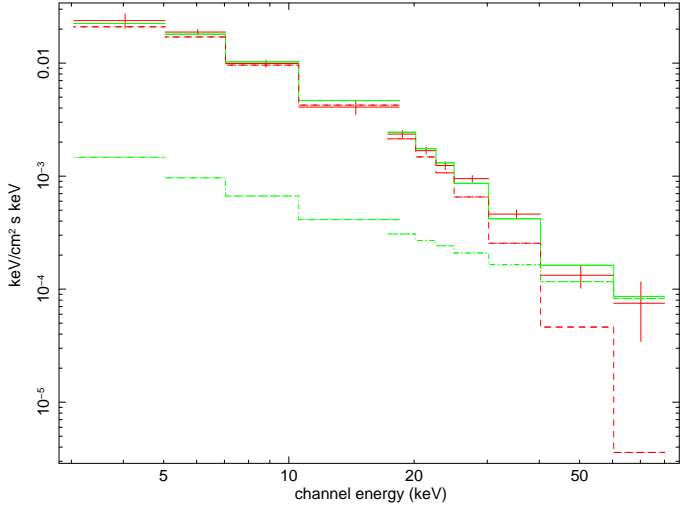}}}
\caption{\intgr\ unfolded spectrum of the Ophiuchus cluster in the 3-80 keV band fitted by a MEKAL+power-law model with fixed photon index $\alpha=2.0$ (green solid line). The red and green dashed lines represent the contribution of the thermal, respectively non-thermal components as a function of energy.}
\label{spec2}
\end{figure}

Fitting the combined JEM-X/ISGRI data with a single-temperature MEKAL model \citep{kaastra}, with the abundance fixed to 0.49 compared to the solar value \citep{mohr} and redshift fixed to $z=0.028$ \citep{johnston}, one finds a bad representation of the data, with a reduced $\chi^2$ exceeding 2.3. Another spectral component is hence needed to get a better fit to the data. Fig. \ref{spec1} shows the combined JEM-X/ISGRI spectrum and the residuals from the thermal model with temperature $kT=8.50$ keV fitted to the spectrum below 20 keV. The model exhibits significant residuals above 20 keV, which could be due to the presence of an additional non-thermal component. Similar to \citet{nevalainen}, we assume that the non-thermal component is a power-law with a fixed photon index $\alpha=2.0$. Fig. \ref{spec2} shows the unfolded spectrum with the thermal model and an additional power-law component fitted simultaneously (Method 1). Table \ref{tabspec} shows the results of the fitting procedure using such a power-law in addition to the thermal emission, with the plasma temperature let free while fitting (Method 1) and fixed to the value found by fitting independantly the 3-20 keV part of the spectrum (Method 2), which is expected to be completely dominated by the thermal component. The \bps\ claim ($2\sigma$ excess at high energies) was obtained by fixing the temperature to the value obtained from the low-energy spectrum, which is similar to Method 2. Therefore, we can see that the clear excess ($6.4\sigma$) detected with this method in \intgr\ data is much more significant than the value obtained by \bps.

\begin{table}
\caption{Results of the fitting procedure for a thermal + power-law model, with abundance and redshift values fixed to the litterature value and photon index fixed to $\alpha=2.0$, similar to the work of \citet{nevalainen}. In Method 1, the temperature and the normalisation of the Bremsstrahlung component are left free while fitting. In Method 2, we used the data in the 3-20 keV band to fix the temperature and normalisation of the thermal component, and let free only the normalization of the non-thermal component. $^a$ $P_{F-test}$ is the null hypothesis probability when adding the non-thermal component given by the F-test. $^b$20-60 keV flux of the non-thermal component, in units of $10^{-12}\mbox{ergs }s^{-1}\mbox{ cm}^{-2}$. The errors are quoted at the $1\sigma$ level. $^c$Confidence level for the detection of the non-thermal component.}
\label{tabspec}
\begin{tabular}{|l|c|c|c|c|c|}
\cline{2-6}
\cline{2-6}
\multicolumn{1}{c}{ }\vline & \, & \, & \, & \, & \,\\
\multicolumn{1}{c}{ }\vline & $\chi^2_{red}$ & ${P_{F-test}}$$^a$ & $kT$ [keV] & Flux$_{HXR}$$^b$ & ${CL_{HXR}\sigma}$$^c$\\
\multicolumn{1}{c}{ }\vline & \, & \, & \, & \, & \,\\
\cline{2-6}
\hline
\, & \, & \, & \, & \, & \,\\
Method 1 & 0.93 & $2\cdot 10^{-4}$ & $8.56_{-0.35}^{+0.37}$ & $10.1\pm2.5$ & 4.0\\
\, & \, & \, & \, & \, & \,\\
Method 2 & 1.05  & $7\cdot10^{-5}$ & $8.50_{-0.45}^{+0.48}$ & $8.2\pm1.3$ & 6.4\\
\, & \, & \, & \, & \, & \,\\
\hline
\end{tabular}
\end{table}

The value found for the plasma temperature is consistent with the \bps\ measurement of $kT=9.1\pm0.6$ keV, but significantly lower than the \emph{ASCA} value of $kT=10.9\pm0.4$ keV. However, the \emph{ASCA}/GIS instrument was sensitive only up to 10 keV, and thus could not detect the cutoff. Because of the sensitivity above 10 keV, the \intgr\ and \bps\ values certainly provide more accurate measurements.

Fixing the thermal component to the value found in the 3-20 keV band, one can try to constrain the acceptable range for the photon index of the non-thermal component. Leaving only the photon index and the normalisation of the power-law free while fitting, one finds $\alpha=1.62_{-0.35}^{+0.32}$. The photon index of 2.0 used to derive the values of the HXR emission is compatible (within $1\sigma$) with this value. With such a hard photon index, it is clear that the excess cannot be explained by the presence of hot gas. Indeed, if instead of a power-law we fit the excess with a second thermal component, we can derive a lower limit to the temperature of the second component, which we find to be $kT> 50$ keV. Together with the temperature map from both JEM-X and \emph{ASCA}, this proves that the excess cannot be explained by the presence of very hot gas.

\subsection{Possible origins of the non-thermal component}
\label{secrad}

Since the hard X-ray flux observed by \intgr\ cannot be explained by thermal Bremsstrahlung, there are two possible origins for the non-thermal flux in the hard X-ray band. Indeed, it can be due either to inverse-Compton (IC) scattering of GeV electrons with the CMB \citep{ensslin}, or to synchrotron emission from another population of very high energy electrons ($E\sim$ PeV, e.g. \citet{timokhin}, \citet{inoue}). To construct our models, we assume that the electron distribution is a power-law with spectral index 2.0 and a high-energy cut-off.

\begin{figure}
\resizebox{\hsize}{!}{\hbox{
\includegraphics{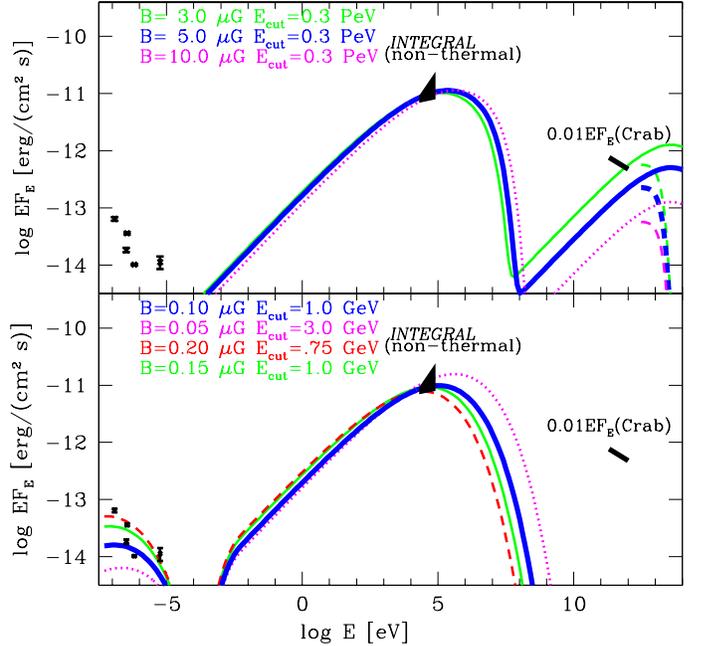}}}
\caption{Models for the spectral energy distribution of the non-thermal emission of the cluster, for a population of $E\sim$ PeV electrons (top) and of $E\sim$ GeV electrons (bottom), for different values of magnetic field and cut-off energy. The solid black line shows the 10 mCrab flux level in the TeV range. The radio data are from \citet{johnston}. The dashed lines at energies $\sim10$ TeV show the attenuation of the IC spectrum due to absorption by the extragalactic infrared and CMB backgrounds.}
\label{radiospec}
\end{figure}

Assuming that the non-thermal component comes from IC scattering of the same electrons that produce radio halos, one can try to constrain the physical parameters needed to explain the radio/hard X-ray flux \citep{thierbach}, in particular the value of the magnetic field in the cluster. Unfortunately, the existing radio data are old, and the size of the region used to extract the radio flux differs for all the measurements, so the shape of the radio spectrum cannot be well constrained. Using the radio data presented in \citet{johnston}, we tried nevertheless to consider jointly the radio and hard X-ray data of the cluster to constrain models for the spectral energy distribution. The bottom panel of Fig. \ref{radiospec} shows the corresponding spectral energy distribution, with models computed for several values of magnetic field $B$ and cut-off energy $E_{cut}$.

Because of the low quality of the radio data, we cannot constrain the properties of the electron population. However, the ratio between the synchrotron and IC components is given by
\begin{equation}\frac{P_{sync}}{P_{IC}}=\frac{u_B}{u_\gamma}=\frac{B^2/8\pi}{u_{\mbox{\tiny{CMB}}}}.\end{equation}
The only unknown factor in this ratio is the magnetic field, so we can use the mean value of the radio data to estimate the value of $B$. Figure \ref{radiospec} shows the spectral models computed for different values of $B$. We can see that the best estimate for the magnetic field is
\begin{equation}B\sim0.1-0.2\,\mu G.\end{equation}
This value is similar to the value obtained from the hard X-ray radiation of the Coma cluster ($B\sim0.15\,\mu G$, \citet{fusco3}), but contradicts the much higher values obtained from Faraday rotation measures in a large sample of radio-emitting clusters \citep{kim}.

If the hard X-ray emission comes from synchrotron radiation from a different population of electrons with highly relativistic velocities, we expect that the peak of the IC emission will be in the TeV domain. The top panel of Fig. \ref{radiospec} shows the corresponding spectral energy distributions for different values of magnetic field and a cut-off energy $E_{\mbox{\tiny{cut}}}=0.3$ PeV. The black line shows the 10 mCrab flux level in the TeV domain. We can see that the magnetic-field values deduced from Faraday rotation measures ($\sim1-5\,\mu G$) predict a TeV flux which is firmly detectable by the present generation of Cherenkov telescopes.

\section{Discussion}
\label{secdisc}

The combined ISGRI/JEM-X spectrum of the cluster exhibits a clear ($6.4\sigma$) excess at high energies. At such significance level, it is now clear that an additional component is needed to explain the relatively high hard X-ray flux observed both by \intgr\ and \bps. Unlike the PDS instrument on board \bps, the ISGRI and JEM-X instruments on board \intgr\ are imaging, which allowed us to study the changes in temperature throughout the detected region, and show that the hard excess cannot be explained by the presence of very hot thermal plasma in some regions of the cluster. Besides, in the large FOV of the PDS instrument, there was no guarantee that the possible excess was not the result of isolated point sources, for instance absorbed active galactic nuclei. The imaging capabilities of \intgr\ allow us to exclude this possibility. Indeed, we find evidence that the source is extended even above 20 keV, and coincides spatially with the X-ray emission. Moreover, high angular resolution X-ray imaging of the cluster with \emph{Chandra} does not reveal the presence of any bright point source below 10 keV in the region detected by \intgr, which shows that contamination of the \intgr\ spectrum by point sources is very unlikely. Therefore, we claim with good confidence that the high-energy tail observed by \intgr\ and \bps\ is certainly due to non-thermal emission from relativistic particles accelerated during a major merging event in the recent past.

Assuming that the non-thermal component is explained by IC scattering from the same electrons that produce the synchrotron radio halo, we used the existing radio data together with the \intgr\ measurement to make a joint model for the synchrotron/IC components, and estimated the magnetic field value to be $B\sim0.1-0.2\,\mu G$. This value is similar to that estimated with the same method for the Coma cluster \citep{fusco3} and with the typical magnetic field values extracted from equipartition assumption \citep{thierbach}, but contradicts the higher values ($B\sim1-5\,\mu G$) derived from Faraday rotation measures (\citet{kim}, \citet{clarke}). To resolve the discrepancy between the Faraday rotation and Inverse-Compton methods, \citet{goldschmidt} suggested that the magnetic field might be decreasing with radius, such that in the outer regions the electron cooling through IC emission would dominate, while in the center, the electrons would radiate more strongly in the radio domain. However, the spatial resolution of \intgr\ allows us to measure the magnetic field within a region of radius less than 6 arcmin from the center of the cluster, and therefore to show that the value of $B$ measured from the HXR radiation does not come from the outer regions of the cluster. More recently, \citet{beck} suggested that the presence of turbulence might introduce a bias in the value of $B$ derived from Faraday rotation measures. From X-ray imaging of the Coma cluster, \citet{schuecker} demonstrated that turbulence is indeed playing an important role in unrelaxed clusters. Although it is much less studied, Ophiuchus is similar to Coma in many aspects, so this explanation could apply as well in our case. In any case, the measurement of the magnetic field through Faraday rotation measures relies on several assumptions, whereas the expected radio/hard X-ray flux correlation provides a direct measurement of the magnetic field, so the lower values obtained in both Ophiuchus and Coma ($B\sim0.1\,\mu G$) certainly provide a better estimate of the cluster magnetic field. Better constraints on the shape of the radio spectrum are therefore important to measure the magnetic field with better accuracy.

On the other hand, if the magnetic field values derived from Faraday rotation measure are correct, the IC emission from the population of electrons that radiate in the radio domain is not sufficient to explain the hard X-ray flux detected by \intgr. This implies the presence of another population of electrons at much higher energies ($E\sim100$ TeV) that would radiate in hard X-rays through synchrotron radiation and would up-scatter the photons of the CMB to produce TeV emission (see the top panel of Fig. \ref{radiospec}). In this case, our models predict a flux in the TeV domain that would be firmly detectable by HESS. Indeed, our lowest curve, which corresponds to the highest allowed value for the magnetic field ($10\,\mu G$), coincides with the sensitivity limit of HESS. Therefore, observations of the cluster in very high energies are crucial to constrain the models of particle acceleration in Ophiuchus in particular and clusters of galaxies in general.

\section{Conclusion}

In this work, we presented the results of broad-band \intgr\ observations of the Ophiuchus cluster in the 3-18 keV band with the JEM-X X-ray monitor and in the 17-80 keV with the ISGRI hard X-ray imager. We have shown that the source is extended for both instruments and that the morphology of the source is consistent with previous studies (Sect. \ref{secima}). We presented a JEM-X hardness-ratio map (7-18 keV to 3-7 keV) and found that the plasma temperature does not vary significantly within the detected region. In Sect. \ref{secspec}, we presented the total \intgr\ spectrum of the cluster in the 3-80 keV band. We explained that the emission cannot be described by a single-temperature MEKAL model, and that an additional spectral component is needed to explain the significant excess found at energies above 20 keV compared to the best fit ($kT=8.5$ keV) for the thermal emission below 20 keV. Consequently, we measure a hard X-ray excess emission from the Ophiuchus cluster, at a much higher confidence level ($6.4\sigma$ for the fixed temperature model) than the \bps\ claim.

With spectral and imaging analysis, we have shown that this excess cannot be explained by the presence of very hot thermal gas or by contamination of the spectrum by point sources. Indeed, fitting the excess with a second thermal component, we found an unrealistic lower limit to the temperature, $kT>50$ keV. Comparing the JEM-X and ISGRI images with X-ray images from higher resolution instruments (\emph{Chandra} and \emph{ASCA}), we found that the morphology of the cluster does not depend on energy, and that no bright point sources are detected in the high-resolutions images, which excludes the possibility that the hard X-ray emission comes from point sources embedded in the cluster. Therefore, we conclude that the high-energy tail is due to non-thermal emission from relativistic electrons, probably accelerated during a merging event in the recent history of the cluster.

To investigate the origin of this non-thermal component, we presented different models for the spectral energy distribution (Sect. \ref{secrad}). If the emission is of inverse-Compton origin from GeV electrons on the CMB, we used the available radio data to construct a joint radio/hard X-ray model for the spectral energy distribution. Comparing the level of the radio and  hard X-ray flux, we were able to estimate the magnetic field in the cluster, which we found to be $B\sim0.1-0.2\,\mu G$. On the other hand, if the hard X-ray emission comes from synchrotron radiation of another population of very high energy electrons, our models predict a detectable flux in the TeV domain. Therefore, observations of the Ophiuchus cluster in the TeV domain, as well as better measurements of the radio spectrum of the cluster, will allow us to constrain the models, and determine if clusters of galaxies are able to accelerate particles up to very high energies.

\begin{acknowledgements}
We would like to thank P. Lubinski for his help on background maps production, I. Telezhinsky for his help on ISGRI imaging analysis, and the JEM-X team for their important support. This work is based on observations with INTEGRAL, an ESA project with instruments and science data centre funded by ESA member states (especially the PI countries: Denmark, France, Germany, Italy, Switzerland, Spain), Czech Republic and Poland, and with the participation of Russia and the USA.
\end{acknowledgements}

\bibliographystyle{aa} 
\bibliography{AA20078853} 
\end{document}